# Teleoperated Robotic Arm Movement Using EMG Signal With Wearable MYO Armband

Hussein F. Hassan, Sadiq J. Abou-Loukh, Ibraheem Kasim Ibraheem *

University of Baghdad, College of Engineering, Department of Electrical Engineering, Al-Jadriyah, 10001 Baghdad, Iraq

husseinalalawy@gmail.com, doctor_sadiq@yahoo.com, ibraheem.i.k@ieee.org

*Corresponding author.

(Teleoperated Robotic Arm Using Myo Armband)

**Abstract**:

The main purpose of this research is to move the robotic arm (5DoF) in real-time, based on the surface Electromyography (sEMG) signals, as obtained from the wireless Myo gesture armband to distinguish seven hand movements. The sEMG signals are biomedical signals that estimate and record the electrical signals produced in muscles through their contraction and relaxation, representing neuromuscular activities. Therefore, controlling the robotic arm via the muscles of the human arm using sEMG signals is considered to be one of the most significant methods. The wireless Myo gesture armband is used to record sEMG signals from the forearm. In order to analyze these signals, the pattern recognition system is employed, which consists of three main parts: segmentation, feature extraction, and classification. Overlap technique is chosen for segmenting part of the signal. Six time domain features (MAV, WL, RMS, AR, ZC, and SSC) are extracted from each segment. The classifiers (SVM, LDA, and KNN) are employed to enable comparison between them in order to obtain optimum accuracy of the system. The results show that the SVM achieves higher system accuracy at 96.57 %, compared to LDA reaching 96.01 %, and 92.67 % accuracy achieved by KNN.

**Introduction**

The electrical signal produced through contraction or relaxation of muscles which are ruled by the nervous system are called Electromyography (EMG) signals. This signal depends on the physiological and anatomical characteristic of muscles and is considered to be a complex signal. The surface electromyography (sEMG) are EMG signals that collect the electrical signals of the muscle activity through placing the electrodes on the surface of the skin. Fig. 1 shows the surface electromyography (sEMG) signals that start with the low amplitude, which changes with muscle contraction activity [1].

Detection of sEMG signals are useful and improve important methodologies in many applications. Such applications are becoming increasingly in demand, in spheres such as biomedical engineering [2], the robotics arm and automation control systems [3,4].

The measurements and precise representations of the sEMG signals depend on the characteristics of the electrodes and their relationship with the skin of the forearm or shoulder, and are affected by the amplifier design, and the transition of the sEMG signals from analogue to digital format [5]. A raw sEMG signal has the maximum voltage of (0-2) mV, and a range of frequency approximately between (0-1000) Hz, but the important frequency that contains useful information lies between (20-500) Hz [6]. The sEMG signals can be acquired by positioning surface electrodes on the arm or the shoulder.

There are two main types of the electrodes that acquire sEMG signals: needle electrodes (inside the skin) and surface electrodes, with no significant variance between them [7]. There are two types of surface electrodes: wired like Myoware muscle sensor or wireless such as Myo gesture control armband. They differ in features, the most important of which is the sampling rate. All these types of electrodes are considered as data acquisition equipment.

Data acquisition represents the first stage in detecting sEMG signals and recording them, processing them through removing noise and unwanted parts of the signal.

The pattern recognition represents the backbone of sEMG signals analysis and processing. In order to move the assisting devices, such as robotic arms or prosthetic limbs, the pattern recognition system is mostly used to obtain gestures of muscle activity. A pattern recognition system consists of three main stages: segmentation, feature extraction selection, and classification [7–9].

In the segmentation stage, the sEMG signals are segmented into slices (windows) or time-slots to extract features for each window, during the muscles' activity. Two very important points in segmentation are the length of segment and segment schema; these need to be specified as the accuracy is affected by them.

The features extraction stage is when suitable features are selected to extract information from each window of the sEMG signals. The features of sEMG signals can be categorized by time domain, frequency domain, and time-frequency transform. Time domain (TD) features extraction is easy to implement and does not require a high computational cost. TD features, such as waveform length (WL), mean absolute value (MAV), autoregressive coefficients (AR), slope sign changes (SSC), and zero crossings (ZC) are extracted from data related to amplitude and frequency of the raw sEMG signals. In contrast, frequency domain (FD) features are extracted widely using power spectral density (PSD). The FD features include median frequency, mean power, peak frequency, maximum amplitude and variance of the central frequency. The Time-Frequency (TF) transform like wavelet transform (WT) and short time Fourier transform (STFT). The TF features require higher computational cost and are more complex compared to TD features [10,11].

In the classification stage, the appropriate classifier algorithm is chosen for the system. The classifier will determine the intended movement depending on the feature-class set previously defined. The classifiers are applied to distinguish different sets of the features. Many techniques are used for the classification purpose, including well-known algorithms, such as linear discriminant analysis (LDA), support vector machines (SVM), k-nearest neighbour (KNN), multilayer perceptron (MLP), artificial neural networks (ANN), multilayer perceptron (MLP), and fuzzy logic (FL) [12].

Researchers and practitioners focus on analysing and processing the sEMG signals to improve the efficiency of information acquired from sEMG signals using different methods. The researchers in [6] reviewed the effect of the electrodes' positions on the efficiency of sEMG signals and selected the best locations and placement for them. Researchers in [13] compared the effect of using Myo gesture armband and Myoware electrodes for data acquisition on accuracy of gestures, although Myo armband has bandwidth limitation. Studies abound in the field of sEMG pattern recognition systems. Whereas the studies in [14,15] competed to find a combination of features with classification related to muscles, to obtain better accuracy for the system, the authors in [11] compared the time domain and frequency domain features. Additionally, the researchers in [16] designed their own classification through fuzzy approach, whilst [17] used a neural network approach and [18] used both neuro-fuzzy classifiers.

The aim of this study is to use Myo gesture control armband as data acquisition of raw sEMG signals rather than Myoware muscle sensor to move the robotic arm (5DoF) in real-time mode, despite its limitation in bandwidth. Overlapped scheme segmentation and MAV, RMS, WL, AR, ZC, and SSC as features extraction, were selected for analysis and processing of raw sEMG signals. This paper also compared three types of classifiers, LDA, KNN, and SVM, to distinguish seven arm movements (wrist right, wrist left, wrist up, wrist down, fist, rest hand and open hand). Finally, the accuracy of the system for each classifier is computed.

The structure of this paper is as follows: Section 2 describes the background of data acquisition and pattern recognition system. Section 3 explains the suggested methodology to control the movement of the robotic arm. The experiments in offline mode and online mode are presented in Section 4. Section 5 explores the simulation results and comparison of the different classifiers. Finally, the conclusion is presented in Section 6.

**Theoretical Background**

The Myo gesture control armband is a wireless wearable technology designed by Thalmic Labs in 2014 with five gestures of hand. It has three parts: gyroscope, accelerometer, and magnetometer. Each part contains the three axes of x, y, z, and all these parts represent the inertial measurement unit (IMU). The Myo gesture armband contains two batteries in different locations; each battery has a capacity of 260 mAh and an operating voltage range of 1.7 to 3.3 volts. The sampling frequency of Myo gesture control armband is 200 Hz. It is used for medical purposes, automation systems applications, and to control the robotic arm. Moreover, this product is supported by SDK kit, which enables communication between the Myo gesture armband and other applications, such as MathWorks' MATLAB[19,20].

In the segmentation section, there are many approaches used to cut off the sEMG signals, such as the adjacent scheme and overlap scheme. The scheme used in this paper is an overlapping scheme. This type of segmentation divides the sEMG signals into equal time slot windows which overlap with each other. Fig. 2 explains slotted windows and how they overlap. Classification decision (D) in overlap technique can be calculated as:

$$D = \frac{1}{2}T_a + \frac{1}{2}T_{new} + \tau \tag{1}$$

where $T_a$ represents the analysis window length, $T_{new}$ represents window increment and $\tau$ represents processing time [10,12].

Furthermore, in this research, six features are extracted for each window: RMS, MAV, SSC, ZC, AR, and WL. All of them are time domain features. In the following, the symbols $f_k$ and N are the sEMG signals in each segment, and the number of samples of the sEMG signals, respectively. The informative data that each feature represents is as follows [10,14]:

1- **Root Mean Square (RMS)**: It has frequency related features, which represent the square root of the mean square of the segment. The mathematical representation of RMS is:

$$RMS = \sqrt{\frac{1}{N}\sum_{K=1}^{N} f_k^2} \qquad (2)$$

2- **Mean Absolute Value (MAV)**: It has amplitude related features, which represent the calculation of the mean absolute value of the segment. The mathematical representation of MAV is:

$$MAV = \frac{1}{N}\sum_{K=1}^{N} |f_k| \qquad (3)$$

3- **Slope Sign Changes (SSC)**: It has frequency related features, which detect changes in the slope sign of the sEMG signal and count them. The mathematical representation of SSC is as follows:

$$SSC = \sum_{k=2}^{n-1} |(f_k - f_{k-1})(f_k - f_{k+1})| \qquad (4)$$

4- **Waveform Length (WL)**: It has amplitude related features, which represent the cumulative length of the sEMG waveform over the time segment. The mathematical representation of WL is as follows:

$$WL = \sum_{k=1}^{N-1} |f_{k+1} - f_k| \qquad (5)$$

5- **Zero Crossings (ZC)**: It has frequency related features, which represent counts of how much signal amplitude cross the zero amplitude over time segment. It is measuring the frequency shift and shows the number of signal sign variations. The mathematical representation of ZC is as follows:

$$ZC = \sum_{k=1}^{N-1} sng(f_k\, f_{k+1}) \cap |f_k - f_{k+1}| \geq \alpha \qquad (6)$$

$$sng(x) = \begin{cases} 1 & \text{if } x \geq 1 \\ 0 & \text{otherwise} \end{cases}$$

where $\alpha$ is threshold.

6- **Auto Regression (AR):** This represents the linear combination of previous windows plus the error term. The mathematical representation of AR is as follows:

$$x_n = \sum_{k=1}^{p} a_k f_{n-k} + e_n \qquad (7)$$

where n = 0, 1... N-1, $a_k$ is the AR model coefficient, $p$ is the AR model's order, $e_n$ is the residual white noise.

Finally, the present study adopts three types of well-known classifier algorithms, which are: support vector machines (SVM), linear discriminant analysis (LDA), and K-nearest neighbour (KNN). The basic concept of SVM is to find a hyper-plane that differentiates the two classes and determines the data on which class area exists. In contrast, KNN depends on the class of (k) point neighbours to determine the class of the data. The LDA model estimates the mean and variance of data for each class.

The key point of the successful system is how to choose the appropriate features to support the classifier. The accuracy of the system depends on the number of correct predictions of the classifier divided by the number of total predictions, measured in percentage [10].

$$Accuracy = \frac{No.\ of\ correct\ predection}{Total\ No.\ of\ predection} \times 100 \qquad (8)$$

**Methodology and Hardware Components**

The system which was designed consists of three main parts: data acquisition, pattern recognition algorithm, and the output; the latter will be discussed in detail later. Fig. 3 shows all parts of the proposed system.

*Data Acquisition and Synchronization*

The Myo gesture control armband is used as data acquisition equipment for recording the training and testing set of sEMG signals. It uses wireless sEMG electrodes (sensors) that surround the forearm and detect the electrical signals of muscle activities. It is a commercially ready product that is suitable for recording sEMG signals and the spatial data related to the trend and motion of the user's forearm.

It consists of eight sEMG steel medical sensors, which are stainless (each sensor represents one channel). It has advantages over other sEMG sensors because no cables are required, the hands can be freely moved, it is easy to wear, relatively cheap, small in size, and lightweight, weighing only 93 g.

The Myo armband has two LEDs; each LED indicates a specific situation. The status LED indicates a Myo gesture armband connected to a computer via Bluetooth and appears in a blue colour. This colour changes to orange when the Myo armband is charged. The logo LED indicates established communication between Myo gesture armband and another application, such as MathWorks' MATLAB. If the LED flashes, it indicates that there is no established connection. In this case, the Myo armband moves a little for reconnection. For people whose hands are not too large, the flex, the adjustable band, is used to reduce the size of Myo gesture armband. Fig. 4 presents the parts of Myo gesture control armband.

To begin, the installation of the Myo connection package is made on a PC to guarantee the connection between Myo gesture armband and a computer via Bluetooth. The TDM64-GCC-4.9.2, MinGW-w64-4.9.2, and Myo-SDK-win-0.9.0 packages are installed to establish the connection between Myo gesture armband and MATLAB R2017. After completing all necessary package installations, the synchronization between Myo armband and MATLAB R2017 is a necessary step before recording sEMG signals. The electrodes (sensors) of the Myo gesture armband senses and records the sEMG signals.

Seven recorded gestures for each training set and testing set are made. Each gesture repeats four times and holds for five seconds. Further, each set starts and ends with a rest hand gesture, thus the length of each set is 145 seconds. Fig. 5 shows the eight channels of the raw sEMG signals that are recorded by Myo gesture armband.

Usually, the collected sEMG signals contain many types of noises due to ambient noise, motion artefact, inherent noise in electronics equipment, and inherent instability of signal. The ambient noise is originated due to radiation of electromagnetic devices, while the motion artefact noise is caused by the interface between the electrode and the skin. The noise generated by electronic devices, as a result, is called the inherent noise. The inherent instability of signal occurs due to the effects of the sEMG signals caused by motor units firing rate for muscles. When using the Myo gesture armband, practically the noise ratio in sEMG signals is low, which does not affect the sEMG data. This feature considers one importance of features in Myo armband. The sEMG signals voltage are very small in nature, about (0-2) mV. The Myo gesture armband amplifies the sEMG signals so that they can be easily handled.

*Pattern recognition system*

The pattern recognition system consists of three stages, as illustrated in Fig. 3. It is used to analyse and process the sEMG signals to distinguish seven hand gestures which are: wrist right, wrist left, wrist up, wrist down, fist, open hand, and rest hand, as shown in Fig. 6.

*Segmentation*

The overlap segmentation technique is used in this study. The window length of sEMG signals plays an important role in classification accuracy and the delay time to process sEMG signals in real-time. When the window length is relatively small, the classification accuracy will be lower, because the information is more distorted. As a consequence, it is difficult to extract useful information from each window. In contrast, if the window length gradually increases, the time delay will also increase, as well as the classification accuracy, to a certain extent, until the continuation in the increase no longer affects the accuracy of the system. Therefore, the trade-off between the accuracy and delay time should be balanced to guarantee a high accuracy and low delay time.

In this work, practically selected window size (win size = 51 samples) and window increment (win_inc = 25 samples) were used for the training set and testing set.

*Feature Extraction*

After segmentation of the sEMG signals into windows of equal size, the types of features to be extracted from each window of sEMG signals were selected, according to those considered the most important in guaranteeing a high accuracy of separation gestures. The information cannot be obtained for features from individual samples of sEMG signals.

In this paper, the time domain features (MAV, RMS, WL, AR, ZC, and SSC) are selected for ease of implementation and no need for high computational resources. These six features were extracted from each window for each channel of the sEMG signals; as a result, the features matrix would be formed. The number of the rows of the matrix represents the total number of windows of the sEMG data, while the number of the columns represents the total number of the features for all channels. Tables 1, 2 and 3 explain the feature effectiveness in each channel for each type of classifier. The less effective features are preferably deleted to decrease computational time, especially in real-time. The SSC feature has less effect on the classification accuracy, as shown in Tables 1, 2, and 3, and was therefore removed from real-time implementation.

*Classification*

Three types of classifiers are used in the system: SVM, LDA, and KNN with k = 3 for comparison to obtain the highest system accuracy among them. The prediction of the classifier would be applied in the next stage as commands for the controller of the robotic arm. If any channel has low efficiency and has little effect on the overall accuracy of the system, it can be removed to gain higher speed computational time, especially in real-time mode, as in the case of channels 2 and 5. Fig. 7 shows the results of applying all selected features extracted with each type of classifier for each channel of sEMG signals.

**System modes**

The system works in two modes: offline mode and online mode (real-time). The offline mode is developed for calculating system accuracy and improving system performance, whilst the online mode is used for moving the robotic arm in real time. These two modes have been implemented through the MATLAB R2017 program.

*Offline mode*

The Myo gesture armband should be worn on the forearm in the same manner, every time it is used to record the training and testing set, so as to guarantee the placements of its sensors in the same position. In order to avoid random readings, this point must be considered during data recording. In this study, the Myo armband is worn on the right forearm so that the status LED is on the sole of the forearm and towards the fingers of the hand, as shown in Fig. 8.

Myo gesture armband was used in this study to collect raw sEMG signals from all channels, during muscle contraction activity. The information acquired or collected by Myo armband is transmitted to the computer via Bluetooth for analysis and processing using MATLAB R2017. The duration of the recording of the training set is 145 seconds, and the same duration for recording the testing set. The window size in both the training and testing set is 51 samples with 25 overlapping samples.

The number of features extracted for each set for all channels is 56, which equals the number of Myo channels (eight channels) multiplied by the number of features (six features: RMS, MAV, SSC, ZC, WL, and AR with order = 2) per window (segment).

Three vectors are calculated for each set; the first vector of the training set is a feature training vector composed of six features extracted from each window for each channel. The second vector is a class training vector representing movements of the training set. The third vector is an index training vector representing the index of starting the movement in the set. A good synchronization between class training vector and index training vector improves movement classification accuracy. For the best synchronization, the programmable model was developed in the MATLAB R2017 to improve the accuracy of the system.

Additionally, the vectors feature testing, class testing, and index testing, calculated from the testing set using the same procedure applied to the training set. The classifier will compare feature testing with feature training in order to expect the predicted class of movements. The accuracy of the system is calculated as equation (8), where the number of correct predictions is the correct output from the compression of predictions class with class testing, while the total prediction is all the expected movements from the classifier. The accuracy of the system using this procedure with the KNN classifier was 92.67 %, while it was 96.01 % with LDA classifier, and 96.57 % with SVM classifier.

*Online mode*

In the real-time mode, the Aideepen ROT3U 5DOF aluminium robotic arm, Due Arduino microcontroller, HP laptop computer, and Myo gesture armband comprise the hardware components of the system. The Aideepen ROT3U robotic arm has a rotation angle of 180 degrees, a height of 42cm (holder closed), clamp maximum opening of 4.5cm, and widest distance of the holder of 10cm. It consists of five servo motors type MG 996R; each servo motor has an operating voltage of 4.8 V to 7.2 V, with running current of 500 mA – 900 mA (6 V), operating speed: 0.17 s/60º (4.8 V), 0.14 s/60º (6 V), and torque: 9.4 kgf·cm (4.8 V), 11 kgf·cm (6 V). Due Arduino specifications are: operating voltage: 3.3V, digital I/O pins: 54 (of which 12 provide PWM output), analogue input pins: 12, DC current for 5V pin: 800 mA, flash memory: 512 KB all available for the user applications, SRAM: 96 KB (two banks: 64KB and 32KB), and clock speed: 84 MHz. HP laptop computer specifications are: processor: Core i5 – 2.3 GHz, RAM: 8 GB. Fig. 9 shows the Due Arduino and robotic arm.

To move the robotic arm as a human arm movement, the Myo armband records the sEMG signals and transmits them to the computer via Bluetooth. The pattern recognition analyses these signals and processes it through MATLAB R2017 software to distinguish these movements. The output of pattern recognition is a vector of predicted movement; each value in this vector represents one gesture. To guarantee a smooth movement of the robotic arm, the majority vote is applied to this vector. Majority vote produces final movements that have been predicted most frequently by the classifier. Due Arduino receives one of the predictions each time to move the robotic arm, as a final result illustrated in Fig. 10.

The response time of robotic arm movement as a result of moving human arm gesture is an important factor in the success measurement of the system. In order to improve the system response, delay time should be reduced. There are many types of delay time in the system, such as Bluetooth transfer delay, computational time of pattern recognition, and Arduino processing time. Bluetooth transfer delay could not be reduced, because it is a standard protocol used by the manufacturing company of Myo gesture armband. The computational time of pattern recognition can be reduced depending on offline mode results; therefore, it is possible to delete factors that do not significantly affect accuracy, such as channels 2 and 5 and the SSC feature, as illustrated in Tables 1, 2, and 3 and Fig. 7, respectively.

The clock speed of Arduino (microcontroller) plays an important role in processing time; therefore, to improve the response time of the system, it is better to select a high clock speed for Arduino. Early experiments used UNO Arduino, which has a clock speed of 16 MHz. Then it was replaced by DUE Arduino that has a clock speed of 84 MHz to reduce delay time and also to improve the response time of the robotic arm. In this mode, the testing set acquired from the Myo armband is directly cut off into segments and the features testing is calculated for these segments. Furthermore, in the classification stage, the classifier expects the predicted vector, by comparing feature testing with feature training, as previously defined. Then the majority vote is applied and passes each value from it to Arduino, which is connected by serial port with the computer to move the robotic arm. Fig. 10 represents the circuit diagram of all parts of the proposed system and how to connect them.

**Results and Discussion**

This section will discuss the results of the experiments and the factors that affect the accuracy of the system. The influential factors are the length of the window, type of features selected and classifier, the position of electrodes or sensors on the forearm (Myo armband wearing style), and the number of channels that are used in the Myo gesture armband.

*Experiment one: effect of window length on system accuracy*

The length of the sEMG signal segment affects the accuracy of classification and the delay time of the system, as explained previously in the segmentation section. In this experiment the window size changes from 25 to 120 samples; during the duration between (25-100) samples, the accuracy of classification increases significantly, but during the duration between (100-120) samples, increasing window size does not lead to a noticeable increase in accuracy. The best results were obtained when the window size is 51 samples, taking into consideration the real-time constraint that should not exceed 300ms for the best response time in online mode. This is shown in Fig. 11, which is obtained from deploying 6 selected features extracted with SVM classifier.

*Experiment two: effect of extracting feature selection on system accuracy*

Selection of features is very important to extract information from sEMG signals because the classifier can distinguish between movements based on extracted information of these features. In this experiment, the accuracy of the system was 96.57 %, 96.01, and 92.67 % for SVM, LDA, and KNN respectively, for six features extracted and eight channels. By observing offline results (Tables 1, 2, and 3), the WL feature had a high effect on the accuracy of the system, but the SSC feature had little effect on the system's accuracy; therefore, this feature can be removed from feature vector set to reduce computational cost in order to reduce processing time. Thus, the online mode depended on five features: AR, ZC, WL, RMS, and MAV. Table 4 explains the accuracy of the system before and after removal of the SSC feature.

*Experiment three: effect of position of electrodes on system accuracy*

The Myo gesture armband is a bracelet shape. Each time the Myo armband is worn differently, the sensor location changes, thus the accuracy of the system also changes. Consequently, the Myo armband should be worn consistently in the same way when recording sEMG signals in the offline mode and in real-time to ensure that positions of electrodes do not change. The Myo armband was worn in several different styles to obtain the best accuracy. The results show that wearing the Myo armband as in Fig. 8 in the offline experiment section achieved the best accuracy.

*Experiment four: effect of number of channels on system accuracy*

Some of the Myo armband sensors will be positioned on muscles that have a little electrical activity which will impact on the value of information collected from that sensor. Therefore, by observing offline results, it was apparent that channels 2 and 5 had the poorest impact on the accuracy of the system. Therefore, to minimize the computational cost, these can be removed from the pattern recognition system. Thus, in online mode, six channels instead of eight channels are considered. Table 5 shows the impact of channels 2 and 5 on the system accuracy.

In summation, this paper's proposed real-time robotic arm control depends on sEMG signals collected by Myo gestures armband model based on six channels (1, 3, 4, 6, 7, and 8) and uses overlap segmentation technique with the window size of 51 samples and 25 overlapping samples. The five features which were extracted from each segment are WL, AR, ZC, MAV, and RMS. SVM classifier was deployed and the Due Arduino connected with (5DoF) robotic arm. The response time was satisfied with the proposed model with an accuracy of over 95 %. Table 6 lists the final results of the proposed model. The results of this study were compared with other studies' results, as shown in Table 7. In this comparison, the movement types and the style of Myo armband wear differ.

**Conclusion**

A new model proposed in this paper to move the robotic arm (5DoF) in real-time depends on recognizing human forearm gestures as based on sEMG signals collected by wireless Myo gesture armband. The model is composed of three stages: signal acquisition, signal pattern recognition system, and controlling the robotic arm. The following conclusions are obtained:

- Using a wearable wireless Myo gesture armband adds flexibility and free movement with a good signal to the system.

- Window size in segmentation affects delay time and system accuracy. There is an inverse relationship between window size and time delay and a fairly direct relationship between it and system accuracy.
- The set of feature extraction of time domain (WL, AR, MAV, RMS, and ZC) produced satisfactory system results for accuracy and time response.
- The SSC time domain feature has little effect on system accuracy while the WL feature has the highest effect.
- An SVM classifier shows good accuracy (95.27 %) in the sEMG signal analysis with time domain feature extraction. LDA, on the other hand, has less accuracy (94.53 %) and KNN has the least accuracy (89.43 %).
- The comparison in Table 7 displays the converged results. It proves that all of the factors mentioned during this study have a clear effect on the ratio of the accuracy of the system, including the length of the window, the number and type of features extracted, and the number of movements classified.

For future research in this field, a recommendation would be to adopt one of the control design methods in [21,22] to add more smoothness to the robotic arm movement.

**Acknowledgments**

I would like to express my thanks and gratitude to Dr. Ali H. Al-Timemy at Al-Khawarizmi College of Engineering, University of Baghdad for his help and continuous support for this research.

**References**

[1] Gheab NH, Saleem SN. Comparison Study of Electromyography Using Wavelet and Neural

Network. Al-Khwarizmi Eng J 2008;4:108–19.

[2]     Wang N, Lao K, Zhang X. Design and Myoelectric Control of an Anthropomorphic Prosthetic Hand. J Bionic Eng 2017;14:47–59. doi:10.1016/S1672-6529(16)60377-3.

[3]     Pham TXN, Hayashi K, Becker-Asano C, Lacher S, Mizuuchi I. Evaluating the Usability and Users' Acceptance of a Kitchen Assistant Robot in Household Environment. 2017 26th IEEE Int. Symp. Robot Hum. Interact. Commun., vol. 2017–Janua, IEEE; 2017, p. 987–92. doi:10.1109/ROMAN.2017.8172423.

[4]     Gonzalo P-J, Holgado-Terriza Juan A. Control of Home Devices Based on Hand Gestures. 2015 IEEE 5th Int. Conf. Consum. Electron. - Berlin, IEEE; 2015, p. 510–4. doi:10.1109/ICCE-Berlin.2015.7391325.

[5]     Day S. Important Factors in Surface EMG Measurement. 2002.

[6]     Ghapanchizadeh H, Ahmad SA, Ishak AJ, Al-Quraishi MS. Review of Surface Electrode Placement for Recording Electromyography Signals. Biomed Res 2017;2017:S1–7.

[7]     ULKIR O, Gokmen G, KAPLANOGLU E. Emg Signal Classification Using Fuzzy Logic. Balk J Electr Comput Eng 2017;5:97–101. doi:10.17694/bajece.337941.

[8]     Samuel OW, Zhou H, Li X, Wang H, Zhang H, Sangaiah AK, et al. Pattern Recognition of Electromyography Signals Based on Novel Time Domain Features for Amputees' Limb Motion Classification. Comput Electr Eng 2018;67:646–55. doi:10.1016/j.compeleceng.2017.04.003.

[9]     Naik GR, Al-Timemy AH, Nguyen HT. Transradial Amputee Gesture Classification Using an Optimal Number of sEMG Sensors: An Approach Using ICA Clustering. IEEE Trans Neural Syst Rehabil Eng 2016;24:837–46. doi:10.1109/TNSRE.2015.2478138.

[10]    Ali AH. An Investigation of Electromyographic (EMG) Control of Dextrous Hand

Prostheses for Transradial Amputees. Plymouth University September, 2013.

[11] Altın C, Er O. Comparison of Different Time and Frequency Domain Feature Extraction Methods on Elbow Gesture's EMG. Eur J Interdiscip Stud 2016;5:35–44.

[12] Nazmi N, Abdul Rahman M, Yamamoto S-I, Ahmad S, Zamzuri H, Mazlan S. A Review of Classification Techniques of EMG Signals During Isotonic and Isometric Contractions. Sensors 2016;16:1304. doi:10.3390/s16081304.

[13] Mendez I, Hansen BW, Grabow CM, Smedegaard EJL, Skogberg NB, Uth XJ, et al. Evaluation of the Myo Armband for the Classification of Hand Motions. 2017 Int. Conf. Rehabil. Robot., IEEE; 2017, p. 1211–4. doi:10.1109/ICORR.2017.8009414.

[14] Huang H, Li T, Bruschini C, Enz C, Koch VM, Justiz J, et al. EMG Pattern Recognition Using Decomposition Techniques for Constructing Multiclass Classifiers. 2016 6th IEEE Int. Conf. Biomed. Robot. Biomechatronics, vol. 2016–July, IEEE; 2016, p. 1296–301. doi:10.1109/BIOROB.2016.7523810.

[15] Al-Timemy AH, Khushaba RN, Bugmann G, Escudero J. Improving the Performance Against Force Variation of EMG Controlled Multifunctional Upper-Limb Prostheses for Transradial Amputees. IEEE Trans Neural Syst Rehabil Eng 2016;24:650–61. doi:10.1109/TNSRE.2015.2445634.

[16] Ajiboye AB, Weir RFH. A Heuristic Fuzzy Logic Approach to EMG Pattern Recognition for Multifunctional Prosthesis Control. IEEE Trans Neural Syst Rehabil Eng 2005;13:280–91. doi:10.1109/TNSRE.2005.847357.

[17] Ahsan MR, Ibrahimy MI, Khalifa OO. Electromygraphy (EMG) Signal Based Hand Gesture Recognition Using Artificial Neural Network (ANN). 2011 4th Int. Conf. Mechatronics, IEEE; 2011, p. 1–6. doi:10.1109/ICOM.2011.5937135.

[18]   Mokhlesabadifarahani B, Gunjan VK. EMG Signals Characterization in Three States of Contraction by Fuzzy Network and Feature Extraction. Singapore: Springer Singapore; 2015. doi:10.1007/978-981-287-320-0.

[19]   Abduo M, Galster M. Myo Gesture Control Armband for Medical Applications. 2015.

[20]   Mannion P. Myo armband : Wearables Design Focuses on Packaging. 2016.

[21]   Ibraheem IK, Abdul-adheem WR. On the Improved Nonlinear Tracking Differentiator Based Nonlinear PID Controller Design. Int J Adv Comput Sci Appl 2016;7:234–41.

[22]   Abdul-adheem WR, Ibraheem IK. Improved Sliding Mode Nonlinear Extended State Observer Based Active Disturbance Rejection Control for Uncertain Systems with Unknown Total Disturbance. Int J Adv Comput Sci Appl 2016;7:80–93. doi:10.14569/IJACSA.2016.071211.

[23]   Benalcazar ME, Jaramillo AG, Jonathan, Zea A, Paez A, Andaluz VH. Hand Gesture Recognition Using Machine Learning and the Myo Armband. 2017 25th Eur. Signal Process. Conf., IEEE; 2017, p. 1040–4. doi:10.23919/EUSIPCO.2017.8081366.

[24]   Mendez I, Pálsdóttir ÁA, Eiríksdóttir DH, Faulkner M, Waris A, Kamavuako EN. Evaluation of Classifiers Performance Using the Myo Armband. Aalborg University, Denmark, 2017.

Table 1: Feature efficiency (%) of the SVM classifier.

| No. of Channel | Features Extracted | | | | | |
|---|---|---|---|---|---|---|
| | RMS | MAV | WL | AR | ZC | SSC |
| Channel 1 | 49.11 | 40.96 | 59.03 | 44.94 | 43.83 | 18.35 |
| Channel 2 | 26.13 | 22.33 | 34.47 | 30.86 | 29.0 | 15.84 |
| Channel 3 | 35.77 | 35.77 | 48.0 | 41.51 | 37.16 | 14.92 |
| Channel 4 | 39.01 | 34.38 | 43.74 | 33.36 | 34.29 | 14.82 |
| Channel 5 | 24.74 | 28.17 | 32.71 | 31.97 | 30.95 | 14.36 |
| Channel 6 | 34.84 | 31.88 | 37.07 | 37.62 | 31.60 | 15.57 |
| Channel 7 | 44.48 | 38.0 | 48.84 | 45.78 | 38.27 | 17.23 |
| Channel 8 | 44.39 | 39.01 | 57.55 | 43.55 | 34.66 | 15.94 |
| All Channels | 90.73 | 88.32 | 96.20 | 87.95 | 89.43 | 20.29 |

Table 2: Feature efficiency (%) of the LDA classifier.

| No. of Channel | Features Extracted | | | | | |
|---|---|---|---|---|---|---|
| | RMS | MAV | WL | AR | ZC | SSC |
| Channel 1 | 54.03 | 48.0 | 53.84 | 43.0 | 38.36 | 15.0 |
| Channel 2 | 25.30 | 20.85 | 35.95 | 32.43 | 28.82 | 15.66 |
| Channel 3 | 30.49 | 37.25 | 42.07 | 38.46 | 33.08 | 14.08 |
| Channel 4 | 40.03 | 39.94 | 41.14 | 34.29 | 28.63 | 14.82 |
| Channel 5 | 25.20 | 33.82 | 35.86 | 29.56 | 31.23 | 14.08 |
| Channel 6 | 34.93 | 35.12 | 43.0 | 36.23 | 29.19 | 15.19 |
| Channel 7 | 43.46 | 31.04 | 47.72 | 44.02 | 37.53 | 16.12 |
| Channel 8 | 34.38 | 31.41 | 43.92 | 40.40 | 33.82 | 15.57 |
| All Channels | 78.59 | 79.33 | 90.91 | 85.82 | 87.95 | 20.01 |

Table 3: Feature efficiency of the KNN classifier.

| No. of Channel | Features Extracted | | | | | |
|---|---|---|---|---|---|---|
| | RMS | MAV | WL | AR | ZC | SSC |
| Channel 1 | 47.08 | 39.85 | 46.89 | 39.85 | 26.22 | 14.92 |
| Channel 2 | 20.29 | 18.62 | 30.13 | 28.17 | 26.32 | 13.99 |
| Channel 3 | 36.42 | 34.84 | 45.13 | 35.95 | 33.73 | 16.03 |
| Channel 4 | 33.82 | 29.93 | 37.44 | 28.73 | 31.32 | 13.90 |
| Channel 5 | 28.35 | 27.71 | 29.0 | 28.08 | 26.32 | 15.01 |
| Channel 6 | 25.39 | 23.63 | 37.16 | 28.54 | 27.06 | 15.19 |
| Channel 7 | 40.68 | 30.76 | 40.96 | 42.63 | 26.04 | 16.86 |
| Channel 8 | 29.10 | 26.87 | 47.26 | 40.50 | 29.0 | 15.38 |
| All Channels | 89.34 | 86.16 | 91.93 | 83.78 | 87.76 | 15.38 |

Table 4: Effects of SSC feature on the system accuracy.

| Classifier | Accuracy with SSC feature | Accuracy without SSC feature |
|---|---|---|
| LDA | 96.01 % | 95.64 % |
| SVM | 96.57 % | 96.38 % |
| KNN | 92.67 % | 93.69 % |

Table 5: Effect of channel 2 and 5 on the system accuracy with six features.

| Classifier | Accuracy with all channels | Accuracy without ch2 & ch5 |
|---|---|---|
| LDA | 96.01 % | 94.25 % |
| SVM | 96.57 % | 94.99% |
| KNN | 92.67 % | 90.54 % |

Table 6: Comparison of system accuracy results of the proposed model.

| Classifier | System Accuracy |
|---|---|
| LDA | 94.53 % |
| SVM | 95.27% |
| KNN | 89.43 % |

Table 7: Comparison results between this study and other studies in offline mode.

| References | This Study | Ref. [2] | Ref. [23] | Ref. [24] |
|---|---|---|---|---|
| No. of gestures | 7 | 8 | 5 | 9 |
| Segmentation (ms) | Overlap Win-size=250 Win-inc=128 | Overlap Win-size=250 Win-inc=70 | Disjoined Win-size=50 | Overlap Win-size=200 Win-inc=50 |
| Feature extraction | MAV, WL, RMS, ZC, SSC, AR 2 | MAV, VAR, AR4, Sample Entropy SE | Dynamic time warping (DTW) | WL, MAV, Willison Amplitude (WAMP), Cardinality (CARD), SSC, ZC |
| Classifier | SVM | LDA | KNN with K = 5 | SVM |
| Accuracy (%) | 96.57 | 97.35 | 86 | 90.43 |

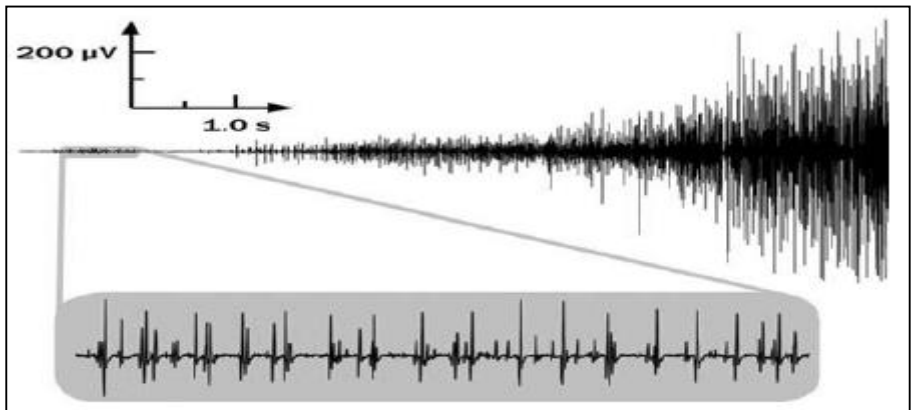

Fig. 1: Estimating the sEMG signal for forearm
muscle via surface electrode [1].

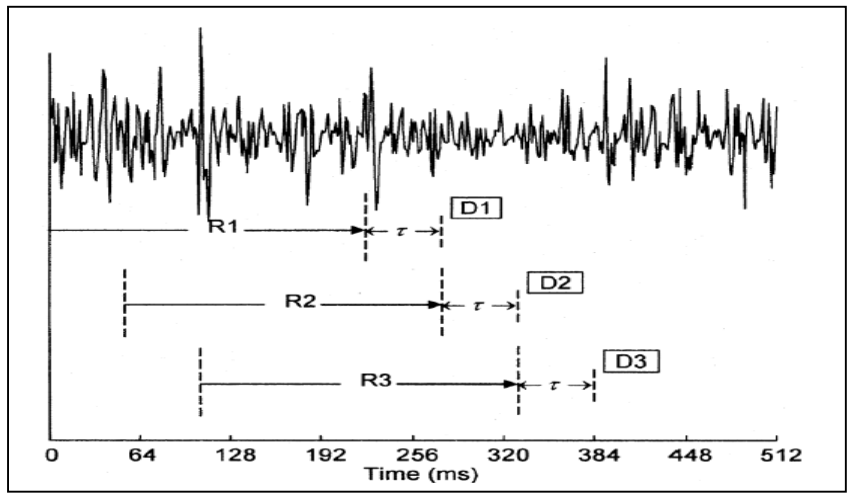

Fig. 2: Overlapping segment technique [10].

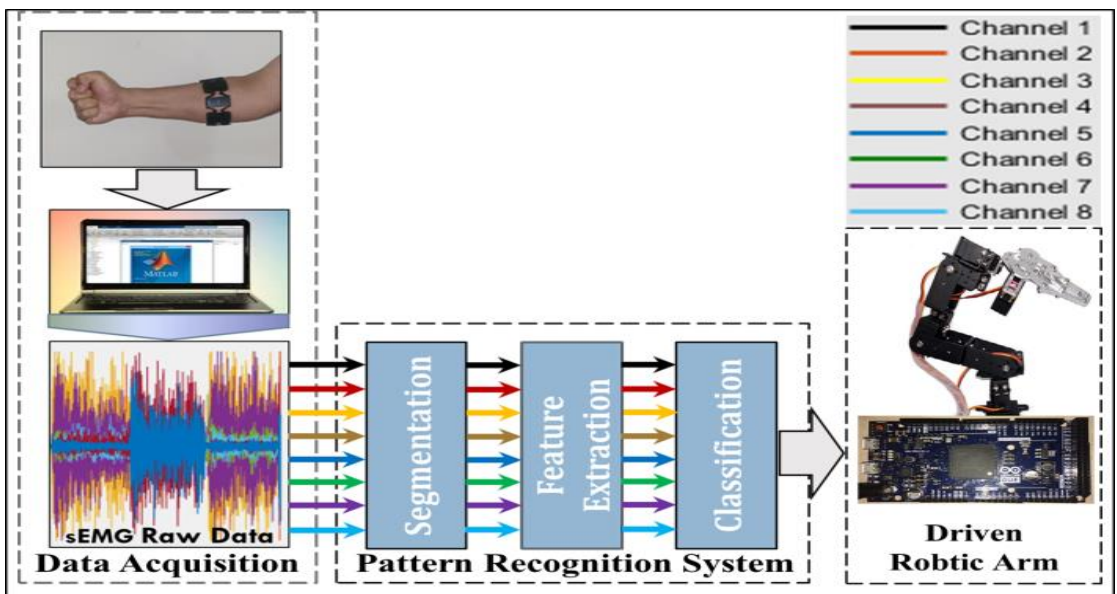

Fig. 3: Block diagram of the overall system proposed.

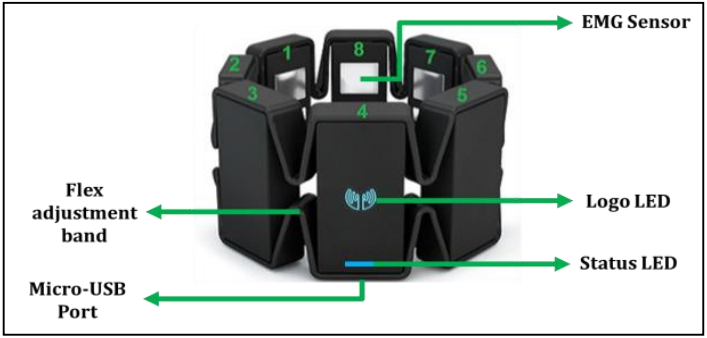

Fig. 4: Myo gesture control armband.

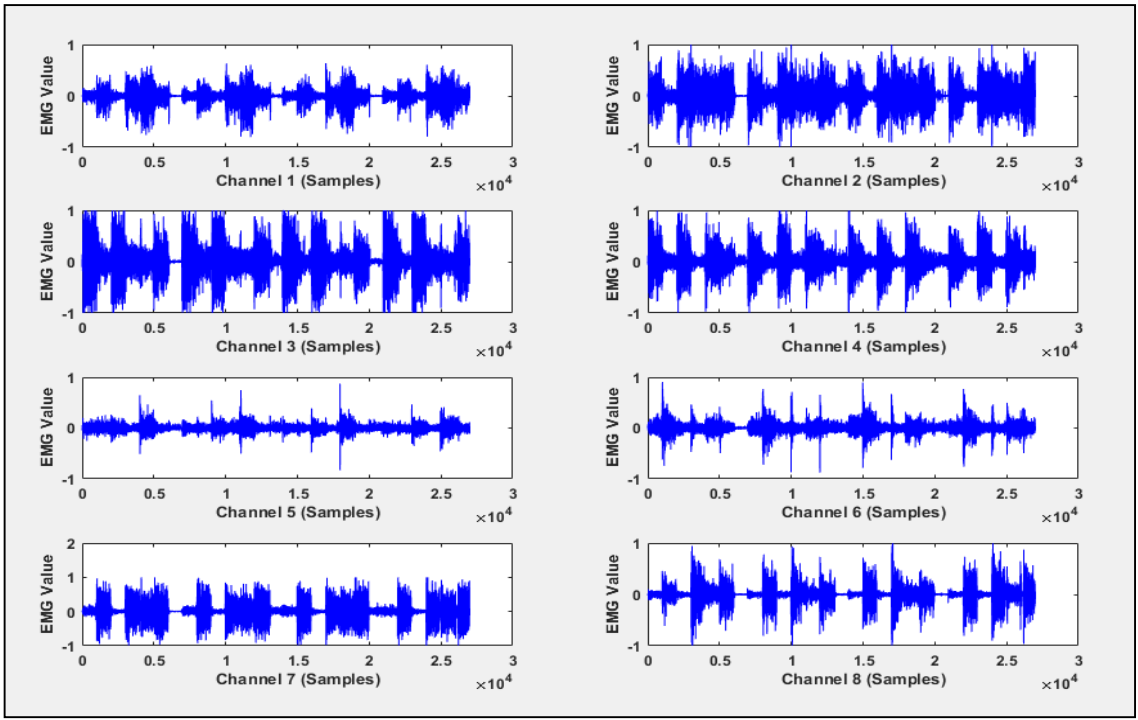

Fig. 5: Raw sEMG signals for eight channels that recorded by Myo armband.

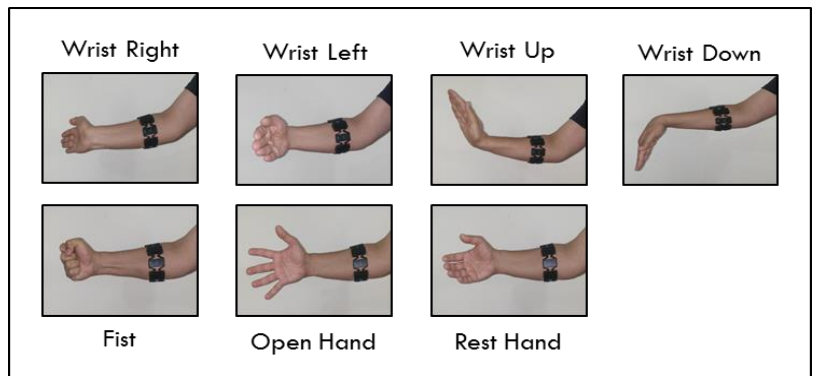

Fig. 6: The seven hand gestures of the proposed system.

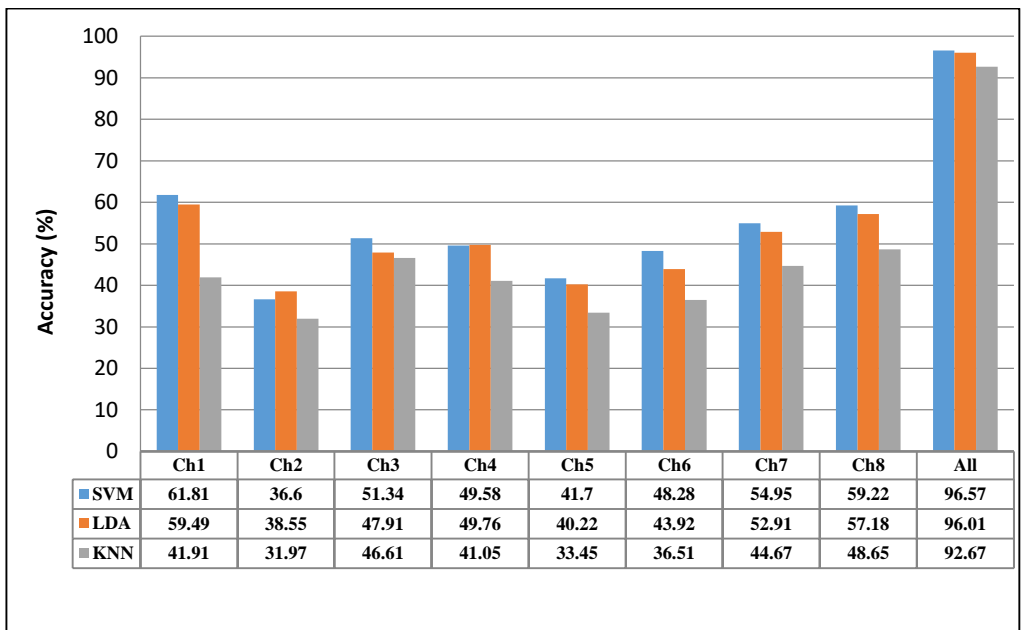

Fig. 7: Classification accuracy (%) of features extraction.

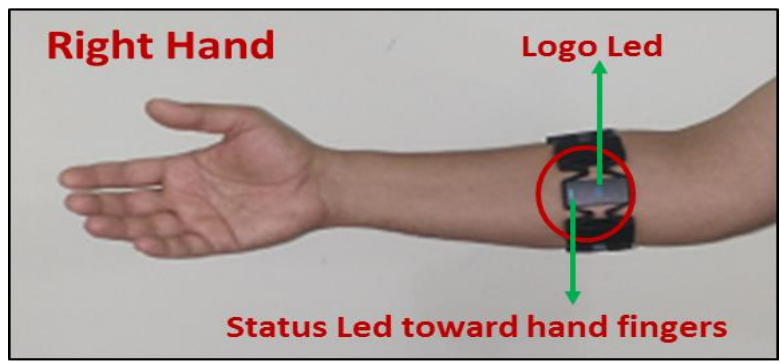

Fig. 8: Manner of wearing Myo gesture armband.

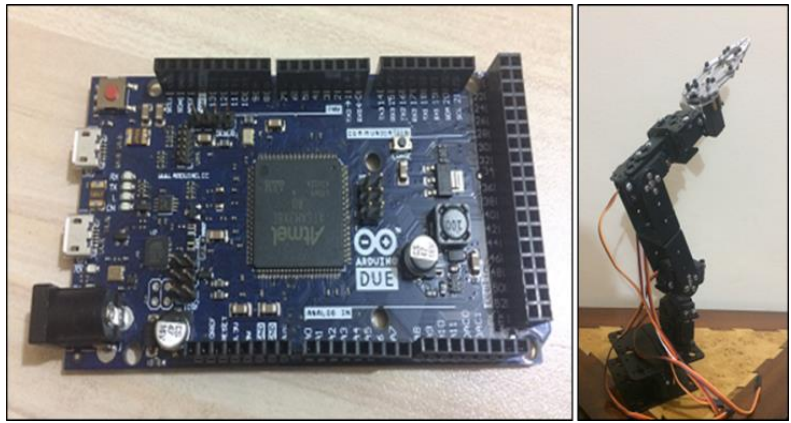

Fig. 9: Due Arduino microcontroller and Aideepen
ROT3U (5DOF) aluminum robot arm.

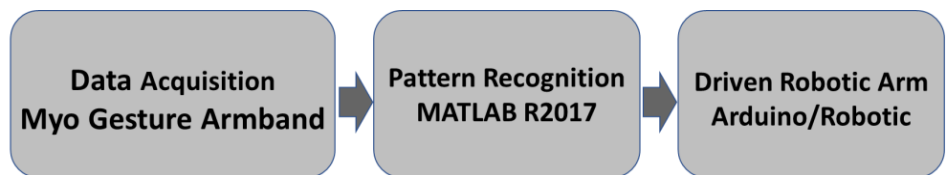

Fig. 10: Block diagram of online mode to move a robotic arm.

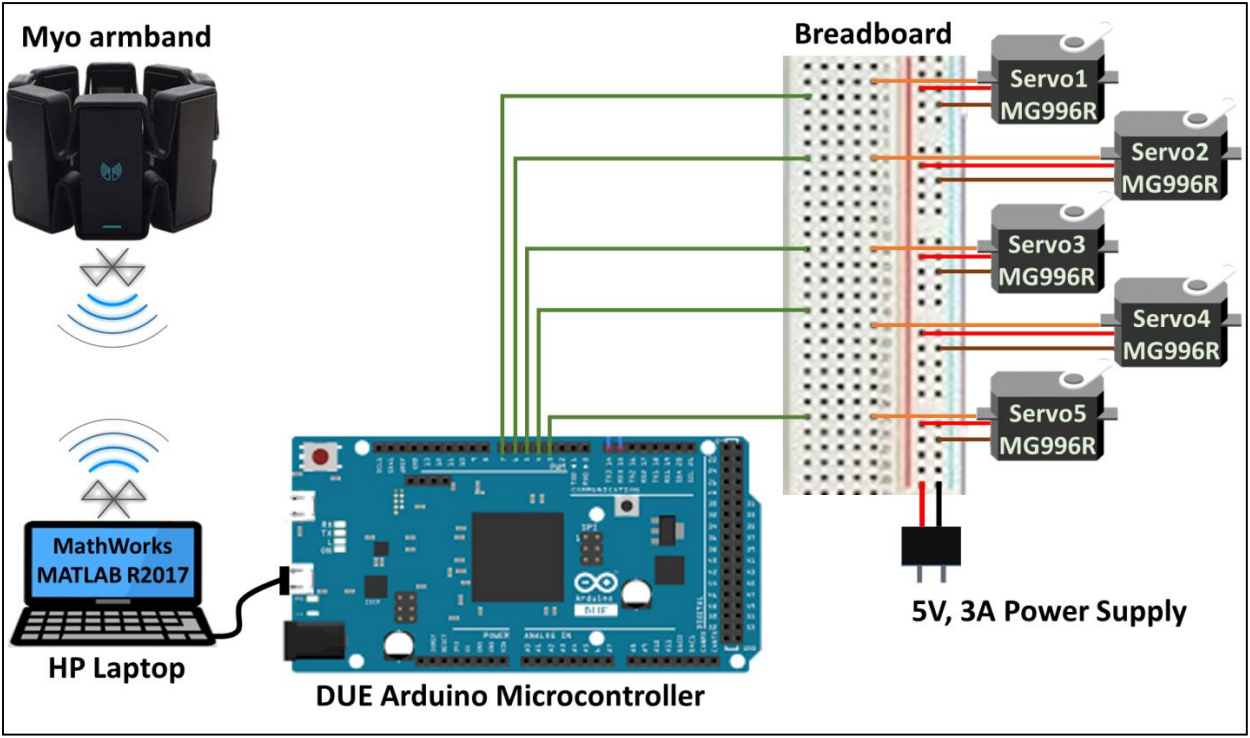
Fig. 11: The circuit diagram of the overall system proposed.

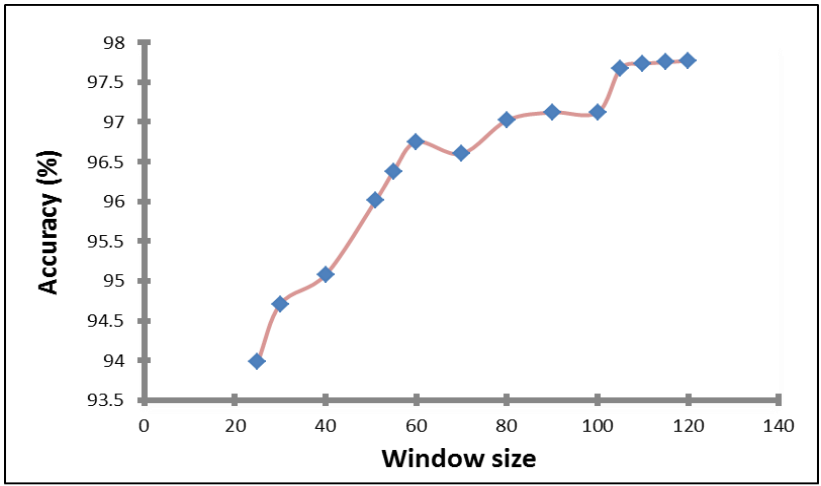

Fig. 12: Effects of window size on system accuracy.